\documentclass[aps,prd,showpacs,twocolumn]{revtex4}

\usepackage{graphicx}
\usepackage{amssymb}
\usepackage{epstopdf}
\usepackage{color}
\usepackage{graphicx}
\usepackage{pdfpages}

\begin{document}

\title{The general relativistic double polytrope for anisotropic matter.
}

\author{G. Abell\'an
\footnote{gabriel.abellan@ciens.ucv.ve}} 
\address{Departamento de F\'isica, Facultad de Ciencias,
Universidad Central de Venezuela, A.P. 47270, Caracas 1041-A, Venezuela.\\}

\author{E. Contreras
	\footnote{econtreras@usfq.edu.ec}}
\address{Departamento de F\'isica, Colegio de Ciencias e Ingenier\'ia, Universidad San Francisco de Quito, Quito, Ecuador.\\}

\author{E. Fuenmayor
	\footnote{ernesto.fuenmayor@ciens.ucv.ve}} \address{Departamento de F\'isica, Facultad de Ciencias,
Universidad Central de Venezuela, A.P. 47270, Caracas 1041-A, Venezuela.\\}

\author{L. Herrera
	\footnote{lherrera@usal.es}} \address{Instituto Universitario de F\'isica Fundamental y Matem\'aticas,
Universidad de Salamanca, Salamanca 37007, Spain\\}

\begin{abstract}
A general formalism recently proposed to study Newtonian polytropes for anisotropic fluids \cite{abellan20} is here extended to the relativistic regime. Thus, it is assumed that  a polytropic equation of state is satisfied by, both, the radial and the trangential pressures of the fluid. Doing so  the generalized Lane--Emden equations are obtained and  solved. Some specific models are obtained, and their physical properties are discussed.
\end{abstract}

\maketitle

\section{Introduction}\label{intro}
In a recent paper an approach for the study of Newtonian polytropes for anisotropic fluids has been proposed \cite{abellan20}. The main idea underlying such an approach consists in complementing the general treatment of Newtonian polytropes  for anisotropic matter described in \cite{LHN}  with the additional assumption that both principal stresses satisfy a polytropic equations of state. Doing so we were able to solve the ensuing Lane--Emden equations, for any set of the parameters of the problem.

Polytropic  equations of state have a long and a venerable history in astrophysics (see \cite{1pol,3pol} and references therein), and have been extensively  used to study the stellar structure under a variety of circumstances. 

For a fluid with isotropic pressure, the theory of polytropes is based on the polytropic equation of state, which in the Newtonian case reads
\begin{equation}P=K\rho_0^{\gamma}=K\rho_0^{1+1/n} ,\label{Pola}\end{equation}
where $P$ and $\rho_0$ denote the isotropic pressure and the  mass (baryonic) density,  respectively. Constants $K$, $\gamma$, and $n$ are usually called  the polytropic constant, polytropic exponent, and polytropic index, respectively. 

Once the equation of state (\ref{Pola}) is assumed, the whole system is described by an equation (Lane--Emden) that may be solved for any set of the parameters of the theory.

However  we know nowaday that, on the one hand,  the pressure isotropy may be a too stringent condition, and  on the other hand that pressure anisotropy is produced by many different physical phenomena, of the kind one expects to be present in very compact objects (see \cite{herrera97,LHN,abellan20} and references therein). Besides, as it has been recently proved, the isotropic pressure condition becomes  unstable by the presence of physical factors such as dissipation, energy density inhomogeneity and shear \cite{LHP}. These facts  explain the renewed interest in the study of fluids not satisfying the isotropic pressure condition, and justify our interest to extend the theory of polytropes to anisotropic fluids.

Thus, if we assume the fluid pressure to be anisotropic, the two principal stresses (say $P_r$ and $P_\bot$) are unequal and the Newtonian polytrope is characterized by the equation:

\begin{equation}P_r=K\rho_0^{\gamma}=K\rho_0^{1+1/n} .\label{Pol}\end{equation}

In this case there is an additional degree of freedom and therefore a single equation of state is not enough to integrate the Lane--Emden equation. In order to overcome this  underdetermination of the problem, we have assumed in \cite{abellan20}, that $P_\bot$ also satisfies a polytropic equation of state.

All  the above concerns the theory of Newtonian polytropes which applies for self--gravitating objects with a degree of compactness of the order of (or lower than) the corresponding to a  white dwarf, for which Newtonian gravity is enough to describe the gravitational interaction. However for more compact objects (e.g. neutron stars, super--Chandrasekhar white dwarfs) we have to resort to general relativity.

For all the reasons above, we endeavour in this work to extend the approach developed in \cite{abellan20} to the relativistic regime. 

General relativistic polytropes  have been extensively studied in the past (see \cite{4a,4b,4c,5,6,7,11,12,13,herrera2013,prnh,pr6,pr9,pr10,pr1,pr7,pr8,pr3,pr5,pr2,pr4} and references therein), in particular a comprehensive framework to describe  general relativistic polytropes for anisotropic fluids has been presented in \cite{herrera2013}. 

As it happens in the Newtonian case, the fact that the principal stresses are unequal, produces in the relativistic case too an underdetermination of the problem, requiring  to impose an additional condition.

Here we propose to follow the same strategy as in \cite{abellan20}, i.e. we shall assume that both the radial ($P_r$) and tangential ($P_\bot$) pressures 
satisfy a polytropic equation of state, i.e. 
\begin{eqnarray}
P_{r}&=&K_{r}\rho^{\gamma_{r}}=K_{r}\rho^{1+1/n_{r}},\label{pr1}\\
P_{\perp}&=&K_{\perp}\rho^{\gamma_{\perp}}=K_{\perp}\rho^{1+1/n_{\perp}}\label{perp1},
\end{eqnarray}
where $\rho$ denotes the energy density.
An important remark is in order at this point:  when considering the polytropic equation of state within the context of general relativity, two different  possibilities arise, leading to the same equation (\ref{Pol}) in the Newtonian limit;  these are (\ref{pr1}) and 
\begin{equation}
P_{r}=K_{r}\rho_{0}^{\gamma_{r}}=K_{r}\rho_{0}^{1+1/n_{r}}\label{pr2}.
\end{equation}
Since we are assuming that the tangential pressure $P_\bot$ also satisfies a polytropic equation of state, it implies that we have  four possible cases leading to the same Newtonian limit. The general treatment is very similar for all these cases and therefore, for simplicity, we shall restrict here to the case described by (\ref{pr1}, \ref{perp1}).

The manuscript is organized as follows. In the next section we introduce the main equations and conventions, and briefly review the main aspects of anisotropic polytropes. Next we introduce the polytropic equation of state for both pressures in section \ref{modeling} and discuss the numerical results. The last section is devoted to final remarks and conclusions.

\section{The polytrope for anisotropic fluid}\label{polyaniso}
\subsection{The field equations and conventions}
We shall  consider  a static, spherically symmetric distribution of an  anisotropic fluid bounded by a surface $\Sigma$.  In Schwarzschild--like coordinates, the metric is parametrized as
\begin{eqnarray}\label{metric}
ds^{2}=e^{\nu}dt^{2}-e^{\lambda}dr^{2}-r^{2}(
d\theta^{2}+\sin^{2}d\phi^{2}),
\end{eqnarray}
where $\nu$ and $\lambda$ are functions of $r$. 

The matter content of the sphere is described by the energy--momentum tensor 
\begin{eqnarray}
T_{\mu\nu}=(\rho+P_{\perp})u_{\mu}u_{\nu}-P_{\perp}g_{\mu\nu}+(P_{r}-P_{\perp})s_{\mu}s_{\nu},
\end{eqnarray}
where, 
\begin{eqnarray}
u^{\mu}=(e^{-\nu/2},0,0,0),
\end{eqnarray}
is the four velocity of the fluid, $s^{\mu}$ is defined as
\begin{eqnarray}
s^{\mu}=(0,e^{-\lambda},0,0),
\end{eqnarray}
with the properties $s^{\mu}u_{\mu}=0$, $s^{\mu}s_{\mu}=-1$.  Notice  that
we are assuming geometric units $c=G=1$.

The metric (\ref{metric}), has to satisfy the Einstein field equations, which are given by
\begin{eqnarray}
\rho&=&-\frac{1}{8\pi}\bigg[-\frac{1}{r^{2}}+e^{-\lambda}\left(\frac{1}{r^{2}}-\frac{\lambda'}{r}\right) \bigg],\label{ee1}\\
P_{r}&=&-\frac{1}{8\pi}\bigg[\frac{1}{r^{2}}-e^{-\lambda}\left(
\frac{1}{r^{2}}+\frac{\nu'}{r}\right)\bigg],\label{ee2}
\end{eqnarray}
\begin{equation}
P_{\perp}=\frac{1}{8\pi}\bigg[ \frac{e^{-\lambda}}{4}
\left(2\nu'' +\nu'^{2}-\lambda'\nu'+2\frac{\nu'-\lambda'}{r}
\right)\bigg]\label{ee3},
\end{equation}
where primes denote derivative with respect to $r$.

Outside of
the fluid distribution the spacetime is given by the Schwarzschild exterior solution, namely
\begin{eqnarray}
ds^{2}&=&\left(1-\frac{2M}{r}\right)dt^{2}-\left(1-\frac{2M}{r}\right)^{-1}dr^{2}\nonumber\\
&&-r^{2}(d\theta^{2}+\sin^{2}d\phi^{2}).
\end{eqnarray}

Furthermore, we require the continuity of the first and  the second fundamental form across the boundary surface $r=r_{\Sigma}=\rm constant$, which implies,
\begin{eqnarray}
e^{\nu_{\Sigma}}&=&1-\frac{2M}{r_{\Sigma}},\label{nursig}\\
e^{-\lambda_{\Sigma}}&=&1-\frac{2M}{r_{\Sigma}}\label{lamrsig}\\
P_{r_{\Sigma}}&=&0,
\end{eqnarray}
where the subscript $\Sigma$ indicates that the quantity is evaluated at the boundary  surface $\Sigma$.

From the radial component of the conservation law,
\begin{eqnarray}\label{Dtmunu}
\nabla_{\mu}T^{\mu\nu}=0,
\end{eqnarray}
one obtains  the generalized Tolman--Oppenheimer--Volkoff equation for anisotropic matter  which reads,
\begin{eqnarray}\label{TOV}
P_{r}'=-\frac{\nu'}{2}(\rho +P_{r})+\frac{2}{r}(P_{\perp}-P_{r}).
\end{eqnarray}
Alternatively, using
\begin{eqnarray}
\nu'=2\frac{m+4\pi P_{r}r^{3}}{r(r-2m)},
\end{eqnarray}
where the mass function $m$ is as usually   defined as
\begin{eqnarray}
e^{-\lambda}=1-2m/r\label{mf},
\end{eqnarray}
we may rewrite Eq. (\ref{TOV}) in the form
\begin{eqnarray}\label{TOV2}
P_{r}'=-\frac{m+4\pi r^{3} P_{r}}{r(r-2m)}(\rho+P_{r})+\frac{2}{r}\Delta, \label{TOVb}
\end{eqnarray}
where
\begin{eqnarray}
\Delta=P_{\perp}-P_{r},
\end{eqnarray}
measures  the anisotropy of the system.

For the physical variables appearing in (\ref{TOVb}) the following boundary conditions apply
\begin{eqnarray}
 m(0)=0,\qquad m(\Sigma)=M, \qquad P_{r}(r_{\Sigma})=0.
\end{eqnarray}
As already mentioned  in the previous section, in order to integrate (\ref{TOV2}), we shall need an additional condition, besides (\ref{pr1}).  In this work such condition is (\ref{perp1}).

\subsection{Relativistic polytrope for anisotropic fluids}
We shall now expose the basics of the theory of relativistic polytropes for anisotropic matter (for details see \cite{herrera2013}).

The starting assumption is to adopt the polytropic equation of state (\ref{pr1}) for the radial pressure, i.e.

\begin{eqnarray}
P_{r}=K\rho^{\gamma_{r}}=
K\rho^{1+\frac{1}{n_{r}}} \label{pr1b}.
\end{eqnarray}

As is well known from the general theory of polytropes, there is a bifurcation at the value $\gamma=1$. Thus, the cases $\gamma=1$ and $\gamma\neq1$ have to be considered separately.

Let us first consider the case $\gamma\neq1$.
Thus, defining the  variable $w$ by
\begin{eqnarray}
\rho=\rho_{c}w^{n_{r}},
\end{eqnarray}
where   $\rho_{c}$  denotes the energy density at the center  (from now on the subscript $c$ indicates that the variable is evaluated at the center), we may rewrite (\ref{pr1b}) as
\begin{eqnarray}\label{Pr}
P_{r}=K\rho_{c}^{\gamma_{r}}w^{n_{r}\gamma_{r}}=P_{rc}w^{1+n_{r}},
\end{eqnarray}
with $P_{rc}=K\rho_{c}^{\gamma_{r}}$. Note that from (\ref{Pr}), we can write
\begin{eqnarray}
P'_{r}=P_{rc}(1+n_{r})w^{n_{r}}w',
\end{eqnarray}
so that (\ref{TOV}) can be written as
\begin{eqnarray}
2P_{rc}(1+n_{r})w'+(P_{rc}w+\rho_{c})\nu'-4\frac{\Delta}{r w^{n_{r}}}=0.
\end{eqnarray}
Next, dividing by $\rho_{c}$ and defining $q_{c}=\frac{P_{rc}}{\rho_{c}}$, we obtain
\begin{eqnarray}
2q_{c}(1+n_{r})w'+(q_{c}w+1)\nu'-4\frac{\Delta}{r\rho_{c}w^{n_{r}}}=0,
\end{eqnarray}
from where
\begin{eqnarray}\label{nup}
\nu'=\frac{4\Delta}{r\rho_{c}w^{n_{r}}(
q_{c}w+1)}-\frac{2q_{c}(1+n_{r})}{q_{c}w+1}w'\label{nup}.
\end{eqnarray}
The  integration of (\ref{nup}) produces
\begin{equation}\label{nur}
\nu=\nu_{c}+\frac{4}{\rho_{c}}\int\limits_{0}^{r}
\frac{\Delta dr}{r(q_{c}w+1)w^{n_{r}}}-2(1+n_{r})\log\left(
\frac{q_{c}w+1}{q_{c}+1}\right).
\end{equation}
To obtain $\nu_{c}$ we can use the boundary condition (\ref{nursig}), from where
\begin{equation}\label{nuc}
\nu_{c}=\log\left(\frac{1-\frac{2M}{r_{\Sigma}}}{(1+q_{c})^{2(1+n_{r})}}\right)-\frac{4}{\rho_{c}}\int\limits_{0}^{r_{\Sigma}}\frac{\Delta dr}{r(q_{c}w+1)w^{n_{r}}}.
\end{equation}
Replacing (\ref{nuc}) in (\ref{nur}) we obtain
\begin{equation}
\nu=\log\left(\frac{1-\frac{2M}{r_{\Sigma}}}{(1+q_{c}w)^{2(1+n_{r})}}\right)-\frac{4}{\rho_{c}}\int\limits_{r}^{r_{\Sigma}}\frac{\Delta dr}{r(q_{c}w+1)w^{n_{r}}},
\end{equation}
whereas replacing (\ref{mf}) and (\ref{nup}) in (\ref{ee2}), produces
\begin{eqnarray}
&&q_{c}w\frac{dm}{dr}+\frac{m}{r}+q_{c}(1+n_{r})\frac{r}{1+q_{c}w}\frac{dw}{dr}\left(1-\frac{2m}{r}\right)\nonumber\\
&&-\frac{2\Delta}{\rho_{c}}\frac{(1-\frac{2m}{r})}{(1+q_{c}w)w^{n_{r}}}=0.\label{le0}
\end{eqnarray}

Let us now introduce the following dimensionless variables
\begin{eqnarray}
\eta&=&\frac{m}{4\pi \rho_{c}\alpha^{3}},\label{d1}\\
r&=&\alpha x,\label{d2}\\
\alpha^{2}&=&
\epsilon\frac{(n_{r}+1)q_{c}}{4\pi \rho_{c}}\label{d3},
\end{eqnarray}
in terms of which  (\ref{le0}) can be written as 
\begin{eqnarray}\label{lemd}
&&\left[\frac{\epsilon x-2q_{c}(1+n_{r})\eta}{1+q_{c}w}\right]
\left[xw'-\frac{2\Delta}{\rho_{c}q_{c}(1+n_{r})w^{n_{r}}}\right]\nonumber\\
&&+\eta+q_{c}xw\eta'=0,
\end{eqnarray}
where $\eta'=x^{2}w^{n_{r}}$
and either $\epsilon=+1$ for $n_{r}>-1$ or $\epsilon=-1$ for $n_{r}<-1$. Please notice that from now on the prime denotes derivative with respect to the variable $x$.

 It is worth noticing that after restoring the speed of light, we have
\begin{eqnarray}
q_{c}=\frac{P_{c}}{\rho_{c}c^{2}},
\end{eqnarray}
implying that  in the Newtonian limit (i.e. $c \to \infty$), we have $q_{c}\to 0$, producing
\begin{eqnarray}\label{le1}
x\left[xw'-\frac{2\Delta}{P_{rc}(1+n_{r})w^{n_{r}}}\right]+\eta =0.
\end{eqnarray}
Then, deriving (\ref{le1}) and using 
$\eta'=x^{2}w^{n_{r}}$ we obtain
\begin{eqnarray}
&&w''+\frac{2}{x}w'+w^{n_{r}}\nonumber\\
&&-\frac{2}{(1+n_{r})P_{rc}w^{n_{r}}x}\left[
\Delta'+\frac{\Delta}{x}-n_{r}\frac{w'}{w}\Delta
\right]=0,\label{le2}
\end{eqnarray}
which is the corresponding equation obtained in   the Newtonian case \cite{abellan20}. 

Let us now consider the case 
$n=\pm \infty$ ($\gamma=1$) which leads to 
\begin{eqnarray}
P_{r}=K \rho.\label{npol1}
\end{eqnarray}
Defining  the dimensionless variable $w$ by
\begin{eqnarray}
\rho=\rho_{c}e^{-w},
\end{eqnarray}
(\ref{npol1}) reads
\begin{eqnarray}
P_{r}&=&K_{r}\rho_{c}e^{-w}=P_{rc}e^{-w}.
\end{eqnarray}

Then, replacing in (\ref{TOV}) we obtain
\begin{eqnarray}
\frac{d\nu}{dr}=\frac{2P_{rc}}{\rho_{c}+P_{rc}}\frac{dw}{dr}
+\frac{4}{(\rho_{c}+P_{rc})r}\Delta e^{w},\label{nuc2}
\end{eqnarray}
or using  (\ref{mf}), (\ref{ee2}) and $q_{c}=P_{rc}/\rho_{c}$ 
\begin{eqnarray}
&&\frac{2e^{w}\Delta}{\rho_{c}(1+q_{c})}\bigg(1-\frac{2m}{r}\bigg)
+\frac{q_{c}r}{1+q_{c}}\bigg(1-\frac{2m}{r}\bigg)\frac{dw}{dr}\nonumber\\
&&\hspace{20pt}-\frac{m}{r}-q_{c}\frac{dm}{dr}=0.
\end{eqnarray}
Finally, in terms of the dimensionless  quantities defined in (\ref{d1}), (\ref{d2}) and (\ref{d3}) the Lane--Emden equation reads
\begin{eqnarray}
&&\left[\frac{2}{q_{c}\rho_{c}}e^{w}\Delta+xw'\right] \left[x-\frac{q_{c}}{1+q_{c}}\eta\right]
\nonumber\\
&&\hspace{40pt}-\eta-q_{c}x \eta'=0\label{case21},
\end{eqnarray}
with $\eta'=x^{2}e^{-w}$.
It is worth noticing that in the limit 
$q_{c}\to 0$ the solution reduces to
\begin{eqnarray}
w''+\frac{2}{x}w'+\frac{2}{P_{rc}x}e^{w}
\left(\Delta'+\frac{\Delta}{x}+\Delta w'\right)=e^{-w},
\end{eqnarray}
which coincides with the Newtonian limit reported in \cite{abellan20}.

\section{The double polytrope}\label{modeling}
As previously commented, in order to solve the problem of the general relativistic polytrope for anisotropic matter,  additional  information (besides (\ref{pr1}))  must be provided. 
In this work this information is supplied by  the assumption that  tangential pressure also satisfies a polytropic equation of state.  Again, due to the bifurcation appearing at $\gamma=1$ , we shall consider separately the cases $\gamma\ne 1$ and $\gamma=1$  mentioned  in the previous section. However, since we have now two polytropic equations of state we shall clearly differentiate two polytropic exponents (indexes) $\gamma_r$, $\gamma_\bot$ ($n_r$, $n_\bot$), one  for each polytrope. Thus,  three possible cases may be considered.
\subsection{Case 1: Both polytropes with $\gamma\ne 1$}

In this subsection we shall assume that $\gamma_r \ne 1$,  $\gamma_\bot\ne 1$, and   the tangential pressure satisfies the polytropic equation of state 
\begin{eqnarray}
P_{\perp}=K_{\perp}\rho^{\gamma_{\perp}},
\end{eqnarray}
whereas the radial pressure satisfies (\ref{pr1}).

Fom the above
\begin{eqnarray}\label{anis}
\Delta=K_{\perp}\rho^{\gamma_{\perp}}-K_{r}\rho^{\gamma_{r}}.
\end{eqnarray}
Introducing $w$ by 
\begin{eqnarray}\label{rnu}
\rho=\rho_{c}w^{n_{r}},
\end{eqnarray}
and replacing (\ref{rnu}) in (\ref{anis}) we may write
\begin{eqnarray}
\Delta=P_{r0}(w^{n_{r}\gamma_{\perp}}-w^{n_{r}\gamma_{r}})=\rho_{c}q_{c}w^{n_{r}}(w^{\theta}-w),
\end{eqnarray}
where $\theta=n_{r}/n_{\perp}$. 
Within this  model the Lane--Emden equation  reads
\begin{eqnarray}
&&\bigg[\frac{x-2q_{c}(1+n_{r})\eta}{1+q_{c}w}\bigg]\bigg[xw'-
\frac{2(w^{\theta}-w)}{1+n_{r}}\bigg]\nonumber\\
&&+\eta+q_{c}xw\eta'=0\label{lem1},
\end{eqnarray}
with $\eta'=x^{2}w^{n_{r}}.$

\begin{figure}[h!]
    \centering
\includegraphics[scale=0.53]{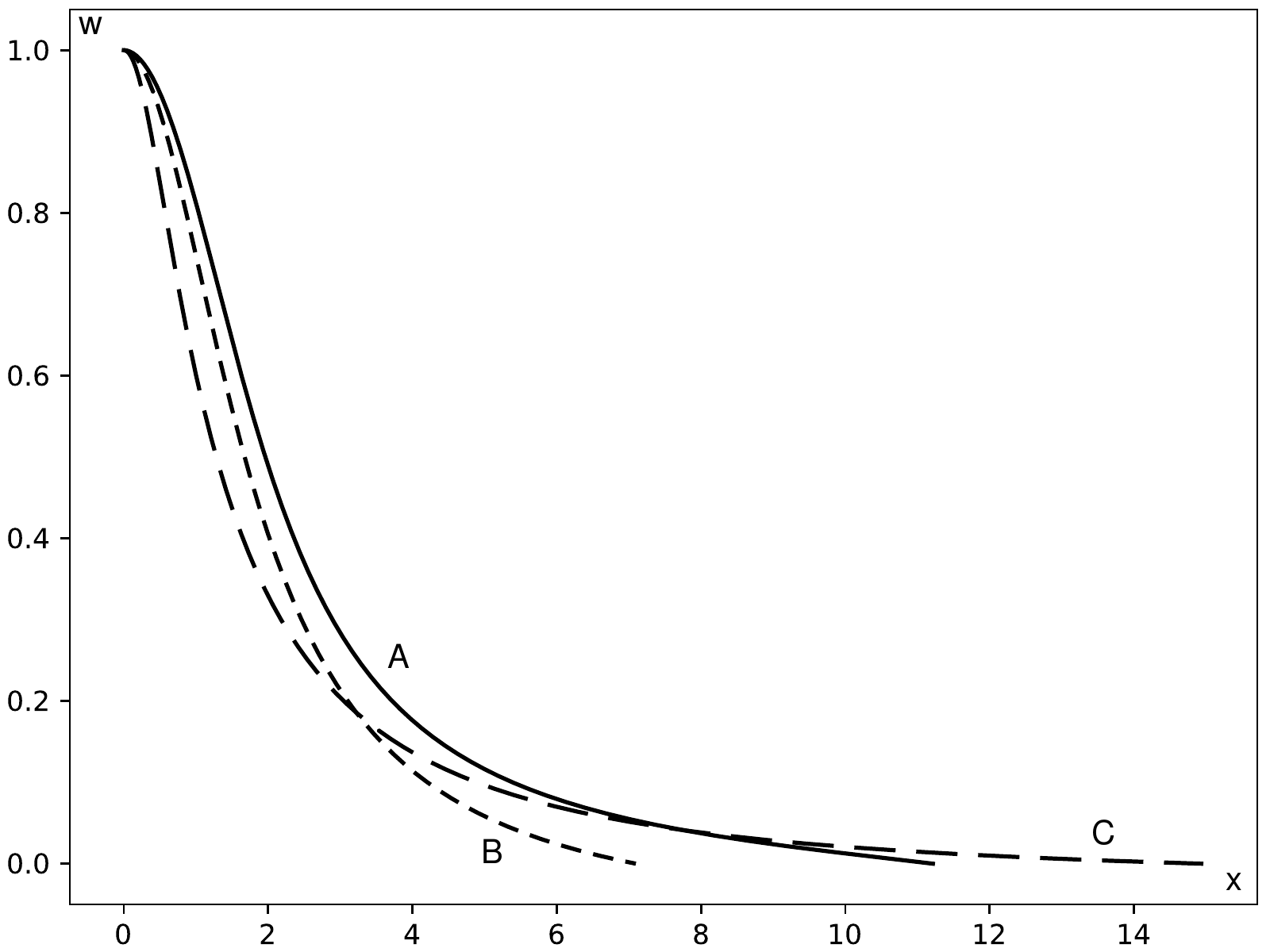}
\caption{Case 1. $w$ as a function of $x$ for $n_{r}=3$, $q_{c}=0.1$ and 
$\theta=0.5$ (short dashed line), $\theta=2$ (long dashed line) and 
$\theta=4$ (solid line)}
\label{case1_w_vs_x}
\end{figure}

In figure \ref{case1_w_vs_x}
it is shown the integration of Eq. (\ref{lem1}) for the values of the parameters indicated in the figure legend.
Note that  $w$ is monotonously decreasing as expected, and the  radial pressure $P_{r}$  vanishes at the surface as required by  the continuity of the second fundamental form.

\begin{figure}[h!]
    \centering
\includegraphics[scale=0.53]{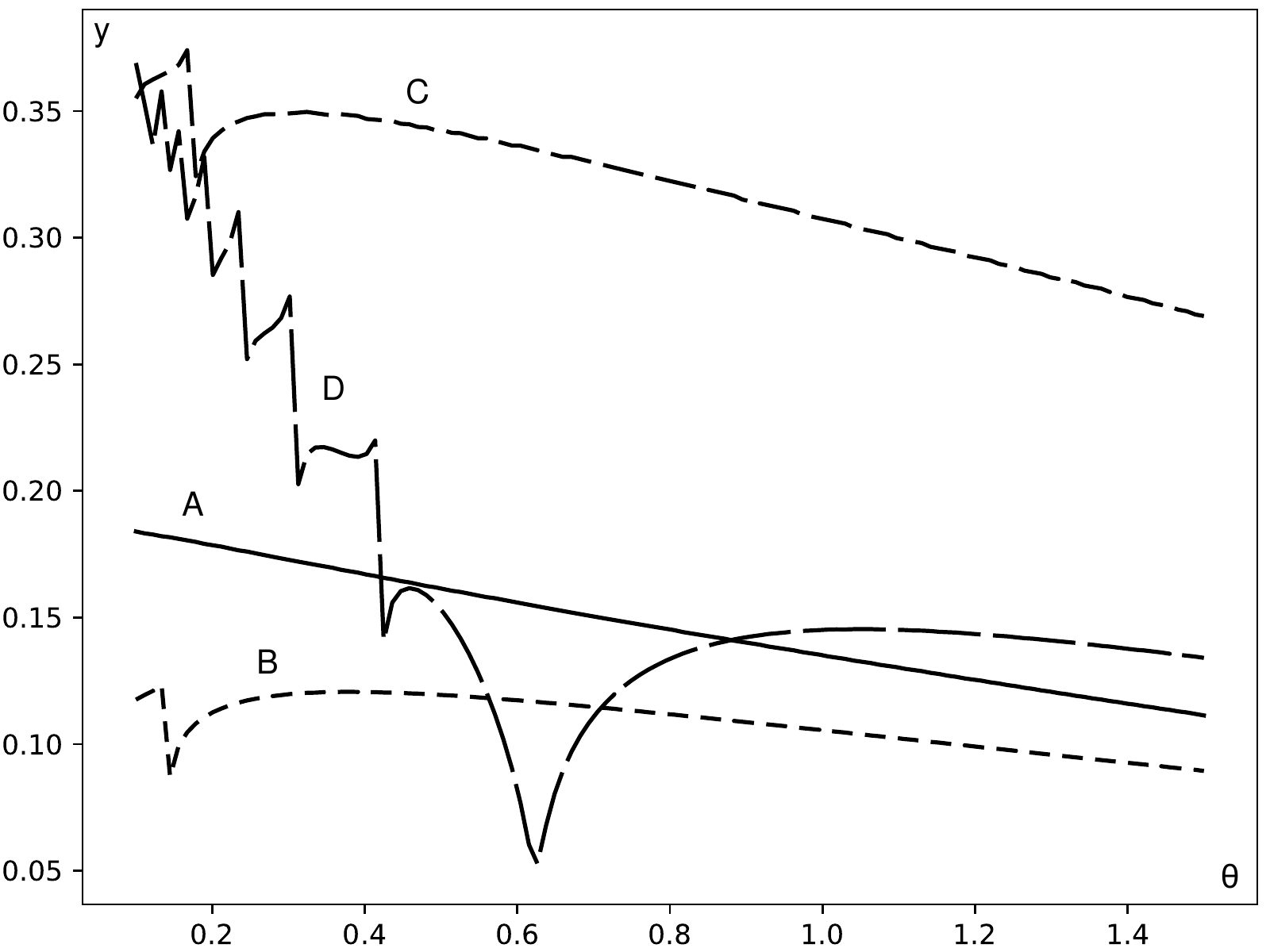}
\caption{Case 1. Surface potential $y$ as a function of the anisotropy parameter $\theta$ for pairs $(q_{c},n_r)$. A: $(0.1,1.0)$ (solid line), B: $(0.1,2.0)$ (short--dashed line), C: $(1.0,1.0)$ (medium--dashed line), D: $(1.0,2.0)$ (long--dashed line).}
\label{comp-case01}
\end{figure}

It will be useful to calculate the Tolman mass, which is a
measure of the active gravitational mass \cite{LHTolman1}, defined by
\begin{eqnarray}\label{mt1}
m_{T}=\frac{1}{2} r^2 e^{\frac{\nu-\lambda}{2}}\nu '.
\end{eqnarray}
Alternatively we can calculate the Tolman mass from the equivalent expression \cite{herrera97}, 
\begin{eqnarray}\label{mt}
m_{T}=e^{\frac{\nu+\lambda}{2}}(m+4\pi r^{3}P_{r}).
\end{eqnarray}
Now, in order to obtain $\nu$ we proceed as follows. First, the TOV
equation (\ref{TOV}) can be written as
\begin{equation}
P_{rc}(1+n_{r})dw=-\frac{1}{2}\rho_{c}(1+q_{c}w)d\nu
+\frac{2P_{rc}}{r}(w^{\theta}-w)dr,
\end{equation}
from where,
\begin{eqnarray}\label{tol}
&&\int\limits_{w(r)}^{w(r_{\Sigma})}\frac{dw}{(1+q_{c}w)}=
-\frac{\rho_{c}}{2P_{rc}}\frac{1}{(1+n_{r})}\int\limits_{\nu(r)}^{\nu(r_{\Sigma})}d\nu\nonumber\\
&&+\frac{2}{1+n_{r}}\int\limits_{r}^{r_{\Sigma}}
\frac{w^{\theta}-w}{(1+q_{c}w)r}dr.
\end{eqnarray}
Next, defining $G(r)$ as
\begin{eqnarray}\label{G}
G(r)\equiv\int\limits_{r}^{r_{\Sigma}}
\frac{w^{\theta}-w}{(1+q_{c}w)r}dr,
\end{eqnarray}
the integration of  Eq. (\ref{tol}) produces
\begin{eqnarray}
-2(1+n_{r})\log (1+q_{c}w)=\nu(r)-\nu(r_{\Sigma})
+4q_{c}G(r),
\end{eqnarray}
from where we obtain
\begin{eqnarray}\label{nut}
e^{\nu}=\frac{1-\frac{2M}{r_{\Sigma}}}{(1+q_{c}w)^{2(1+n_{r})}e^{4q_{c}G(r)}}.
\end{eqnarray}
Finally using  (\ref{mf}) (\ref{d1}), (\ref{d2}), (\ref{d3})  and  (\ref{nut}) in (\ref{mt}) we arrive at
\begin{equation}
\eta_{T}=\left(\frac{1-2y}{1-
2\epsilon q_{c}(1+n_{r})\frac{\eta}{x_{\Sigma}z}}\right)^{1/2} \!\!
\frac{\eta+q_{c}x_{\Sigma}^{3}z^{3} w^{1+n_{r}}}{(1+q_{c}w)^{1+n_{r}}e^{2q_{c}G(z)}},\label{mtc1}
\end{equation}
where 
\begin{eqnarray}
\eta_{T}&=&\frac{m_{T}}{4\pi\rho_{c}\alpha^{3}}\\
y&=&\frac{M}{r_{\Sigma}},\\
z&=&\frac{x}{x_{\Sigma}},
\end{eqnarray}
and
\begin{eqnarray}
G(z)=\int\limits_{z}^{1}\frac{w^{\theta}-w}{(1+q_{c}w)z'}dz'.
\end{eqnarray}

The parameter $y$ (``the surface potential'')  which measures the degree of compactness is plotted in figure \ref{comp-case01} as function of  the anisotropy parameter $\theta$ for different duplets  $(q_c, n_r)$. 

\begin{figure}[h!]
    \centering
\includegraphics[scale=0.53]{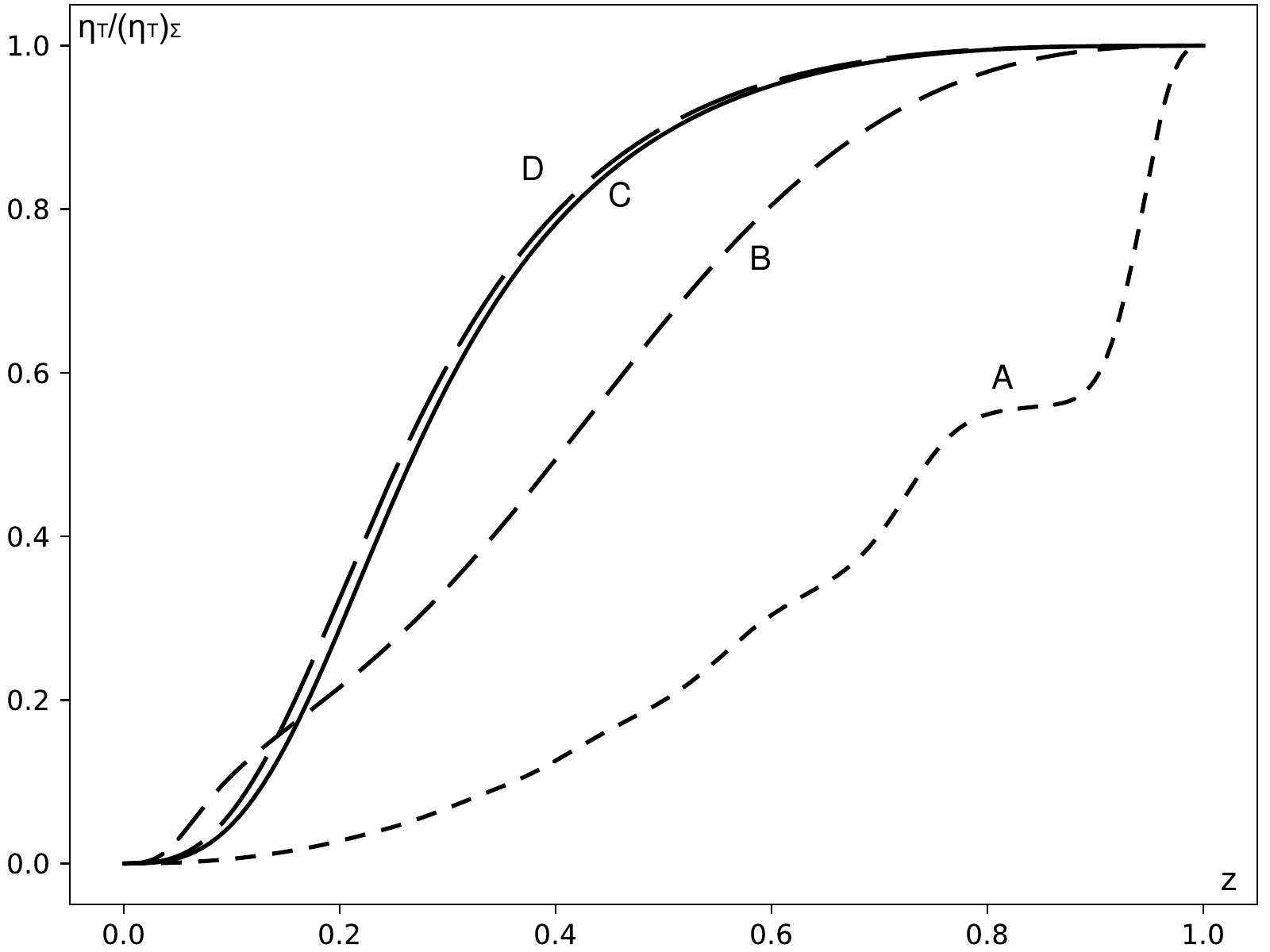}
\caption{Case 1. $\eta_T/(\eta_T)_\Sigma$ as a function of $z$ for $n_r = 2.0$, $q_c = 1.0$ and different values of $y(\theta)$. A: $0.3690(0.1)$ (short--dashed line), B: $0.1550(0.5)$ (medium--dashed line), C: $0.1449(1.0)$ (solid line --isotropic case), D: $0.1341(1.5)$ (long--dashed line). Values of $y$ are read from Fig. \ref{comp-case01}.}
\label{TM-case01-b}
\end{figure}

Figure  \ref{TM-case01-b}  displays the Tolman mass (normalized by the total mass), for the case 1, as function of $z$ for the selection of values of the parameters indicated in the legend. The behaviour of the curves is qualitatively the same for a wide range of values of the parameters. This figure deserves a  detailed analysis.  

Indeed, observe that as we move from the less compact configuration (curve D) to the more compact one (curve A), the Tolman mass tends to concentrate on the outer regions of the sphere, except for curves B, C, D at the innermost regions (see some comments on this point in the last section).  In its turn, more compact configurations correspond to  smaller values of the parameter that measures the anisotropy and smaller values of  the Tolman mass  in the inner regions. In other words, for this case, smaller values of $\theta$ (corresponding to  more compact configurations)  reach stability by reducing  the active gravitational mass in the inner regions. Therefore, it may be inferred from this figure  that more stable configurations correspond to smaller values of  $\theta$ since they  are associated to a sharper reduction of the Tolman mass in the inner regions.

\subsection{Case 2:  $\gamma_r=1$ and $\gamma_\bot\ne 1$}
In this case, besides considering $P_{r}=K_{r}\rho$ with $\rho=\rho_{c}e^{-w}$
we assume
\begin{eqnarray}
P_{\perp}&=&K_{\perp}\rho^{1+\frac{1}{n_{\perp}}},
\end{eqnarray}
from where the anisotropy factor reads
\begin{eqnarray}
\Delta=P_{rc}e^{-w}(e^{-\frac{w}{n_{\perp}}}-1).
\end{eqnarray}

From the above, the Lane--Emden equation  (\ref{case21}) becomes
\begin{eqnarray}
&&\bigg [xw'+2(e^{-\frac{w}{n_{\perp}}}-1)\bigg ]\left(x-\frac{q_{c}}{1+q_{c}}\eta\right)\nonumber\\
&&\hspace{20pt}-\eta-q_{c}x \eta'=0,\label{c2le}
\end{eqnarray}
with $\eta'=x^{2}e^{-w}$.

\begin{figure}[h!]
    \centering
\includegraphics[scale=0.53]{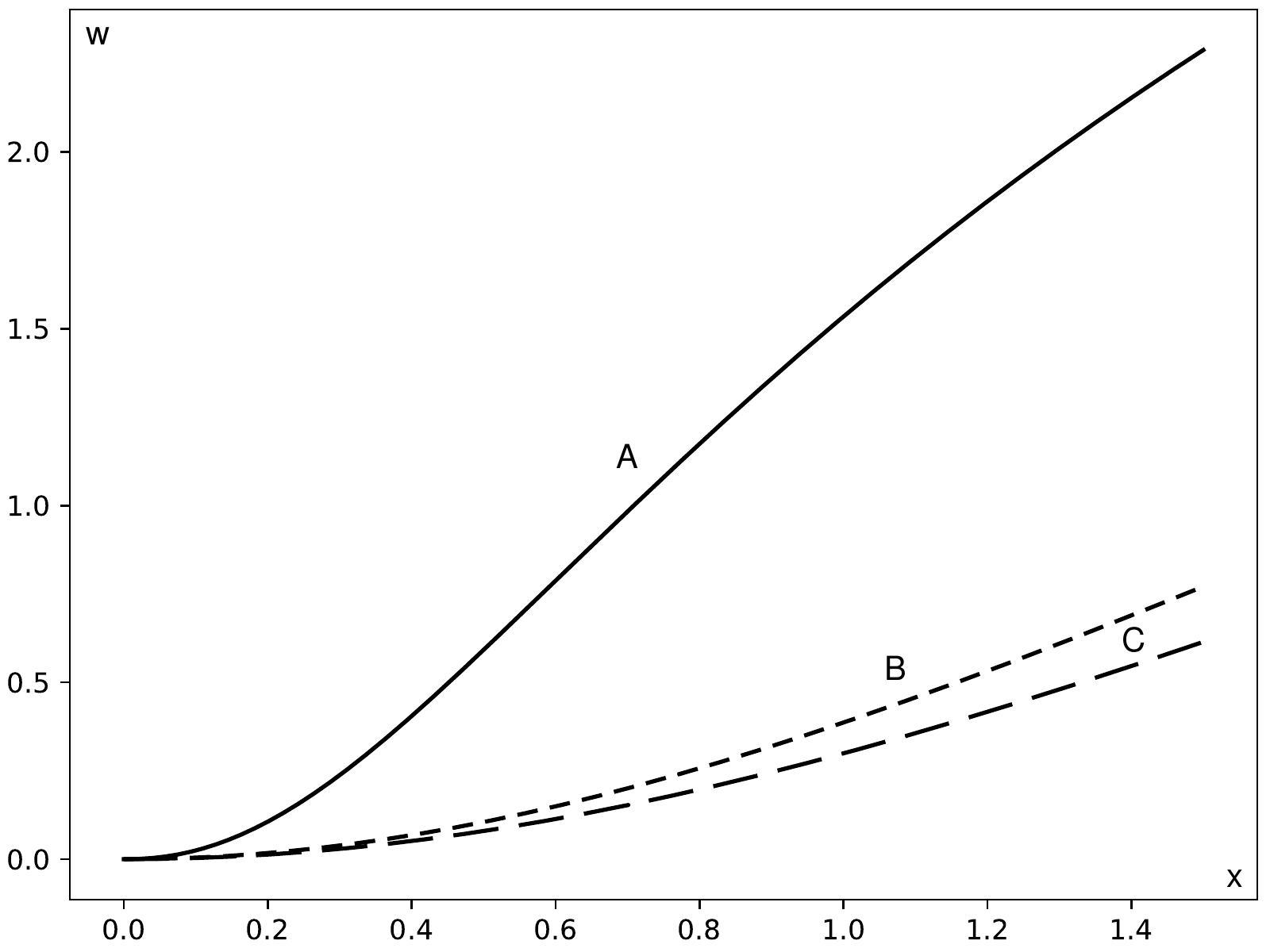}
\caption{Case 2. $w$ as a function of $x$ for $q_{c}=0.1$,  and $n_{\perp}=1$ (short dashed line), $n_{\perp}=2$ (long dashed line) and $n_{\perp}=3$ (solid line)}
\label{case2_w_vs_x}
\end{figure}

In figure \ref{case2_w_vs_x} it is shown the 
integration of Eq. (\ref{c2le})  for  the  values  of  the  parameters  indicated  in the figure legend.

As it is apparent from this figure, these configurations are unbounded and therefore it is meaningless to define a surface potential or the total mass.

\subsection{Case 3:  $\gamma_r\ne 1$  $\gamma_\bot= 1$}
\begin{figure}[h!]
    \centering
\includegraphics[scale=0.53]{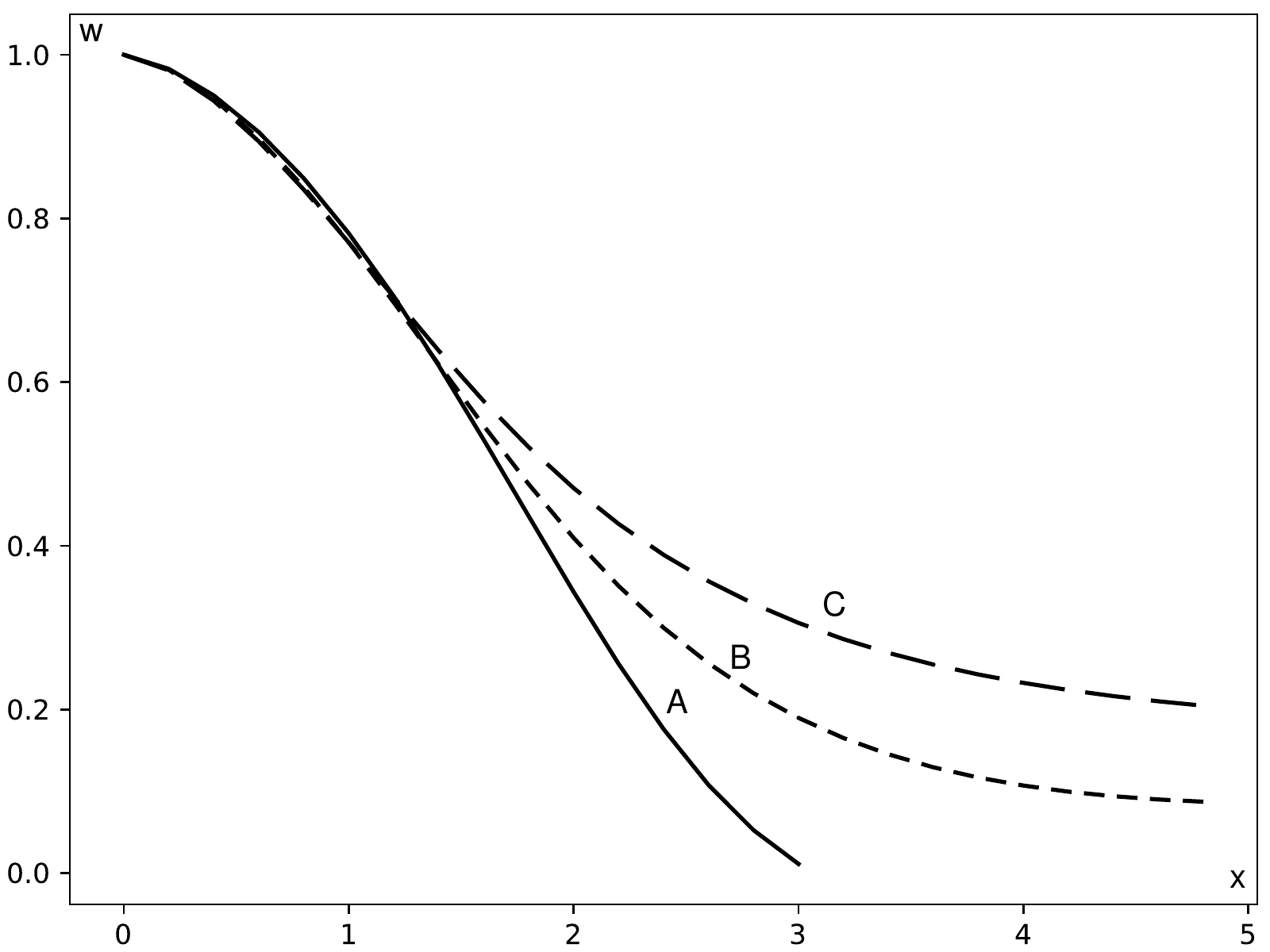}
\caption{Case 3. $w$ as a function of $x$ for $q_{c}=0.1$,  and $n_{r}=1$ (short dotted line), $n_{r}=2$ (long dotted line) and $n_{r}=3$ (solid line)}
\label{case3_w_vs_x}
\end{figure}

In this case we assume
$P_{r}=K_{r}\rho^{1+\frac{1}{n_{r}}}$ with $\rho=\rho_{c}w^{n_{r}}$ and
\begin{eqnarray}
P_{\perp}=K_{\perp}\rho,
\end{eqnarray}
from where the anisotropy reads
\begin{eqnarray}
\Delta=P_{rc}w^{n_{r}}(1-w).
\end{eqnarray}
Then, (\ref{lemd}) becomes
\begin{eqnarray}
&&\bigg[
\frac{\epsilon x-2q_{c}(1+n_{r})\eta}{1+q_{c}w}
\bigg]
\bigg[
xw'-\frac{2(1-w)}{(1+n_{r})}\bigg]\nonumber\\
&&\hspace{40pt}+\eta+q_{c}xw\eta'=0\label{c3le},
\end{eqnarray}
with $\eta'=x^{2}w^{n_{r}}$.
In figure \ref{case3_w_vs_x} it is shown the 
integration of Eq. (\ref{c3le})  for  the  values  of  the  parameters  indicated  in the figure legend.

Next, the TOV equation may be rewritten for this case as,
\begin{equation}
P_{rc}(1+n_{r})dw=-\frac{1}{2}\rho_{c}(1+q_{c}w)d\nu
+\frac{2P_{rc}}{r}(1-w)dr,\,
\end{equation}
and  integrating we obtain
\begin{eqnarray}\label{tol_case3}
&&\int\limits_{w(r)}^{w(r_{\Sigma})}\frac{dw}{(1+q_{c}w)}=
-\frac{\rho_{c}}{2P_{rc}}\frac{1}{(1+n_{r})}\int\limits_{\nu(r)}^{\nu(r_{\Sigma})}d\nu\nonumber\\
&&+\frac{2}{1+n_{r}}\int\limits_{r}^{r_{\Sigma}}
\frac{1-w}{(1+q_{c}w)r}dr.
\end{eqnarray}
\begin{figure}[h!]
    \centering
\includegraphics[scale=0.53]{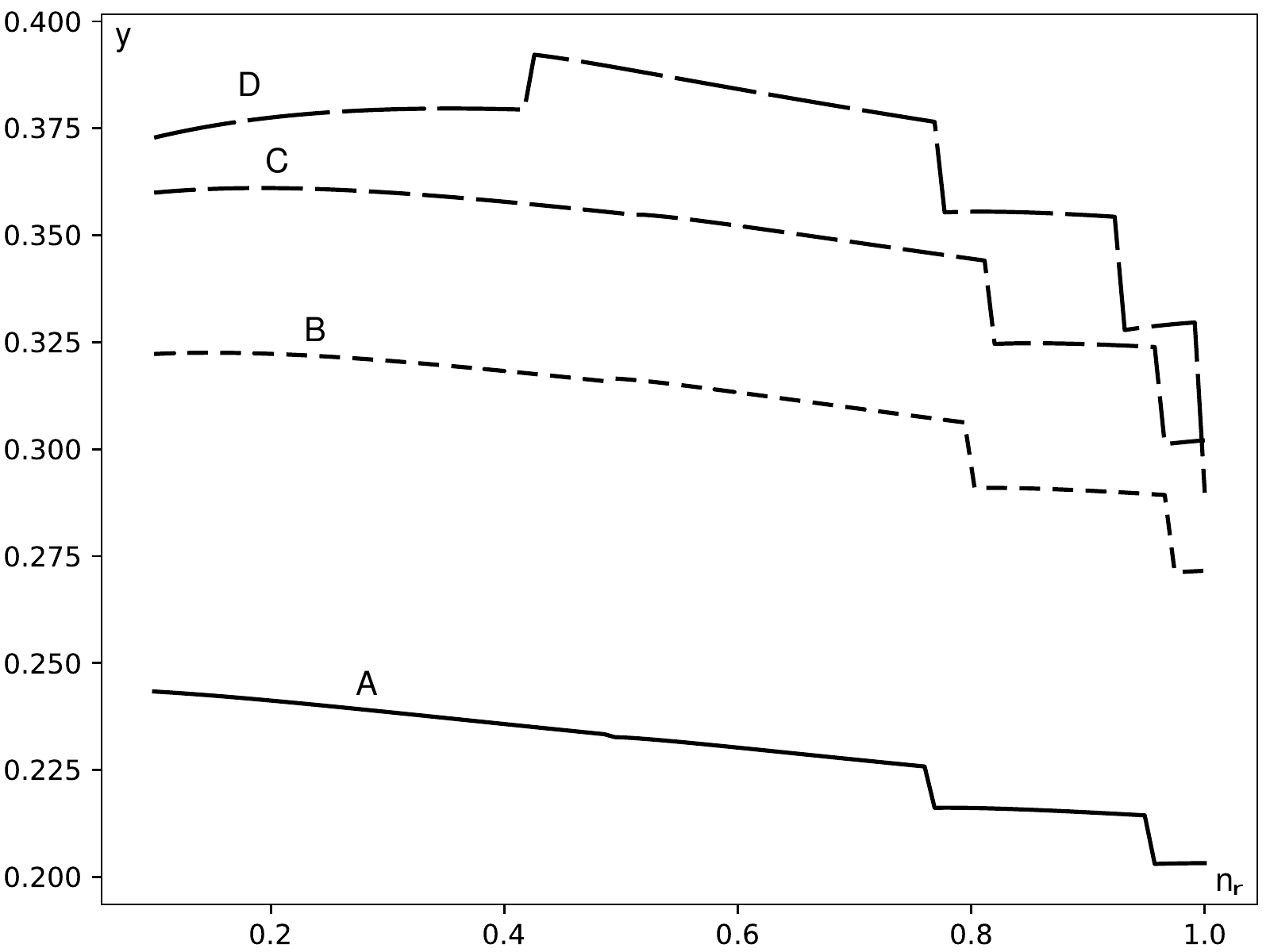}
\caption{Case 3. Surface potential $y$ as a function of  the polytropic index $n_r$ for different $q_{c}$ values. A: $q_c = 0.10$ (solid line), B: $q_c = 0.18$ (short--dashed line), C: $q_c = 0.24$ (medium--dashed line), D: $q_c = 0.30$ (long--dashed line).}
\label{comp-case03}
\end{figure}
The above equation is the equivalent to (\ref{tol}), for this case. They differ in the last term corresponding to the function $G(r)$ . Thus, redefining  the function $G(r)$  for the case 3 by
\begin{eqnarray}\label{G_case3}
G(r)\equiv\int\limits_{r}^{r_{\Sigma}}
\frac{1-w}{(1+q_{c}w)r}dr,\,
\end{eqnarray}
and retracing  the same steps as in the case 1, we found for  the Tolman mass in this case the same expression (\ref{mtc1}), but with a different  function $G(z)$,  which now reads
\begin{eqnarray}
G(z)=\int\limits_{z}^{1}\frac{1-w}{(1+q_{c}w)z'}dz'\,.
\end{eqnarray}
\begin{figure}[h!]
    \centering
\includegraphics[scale=0.53]{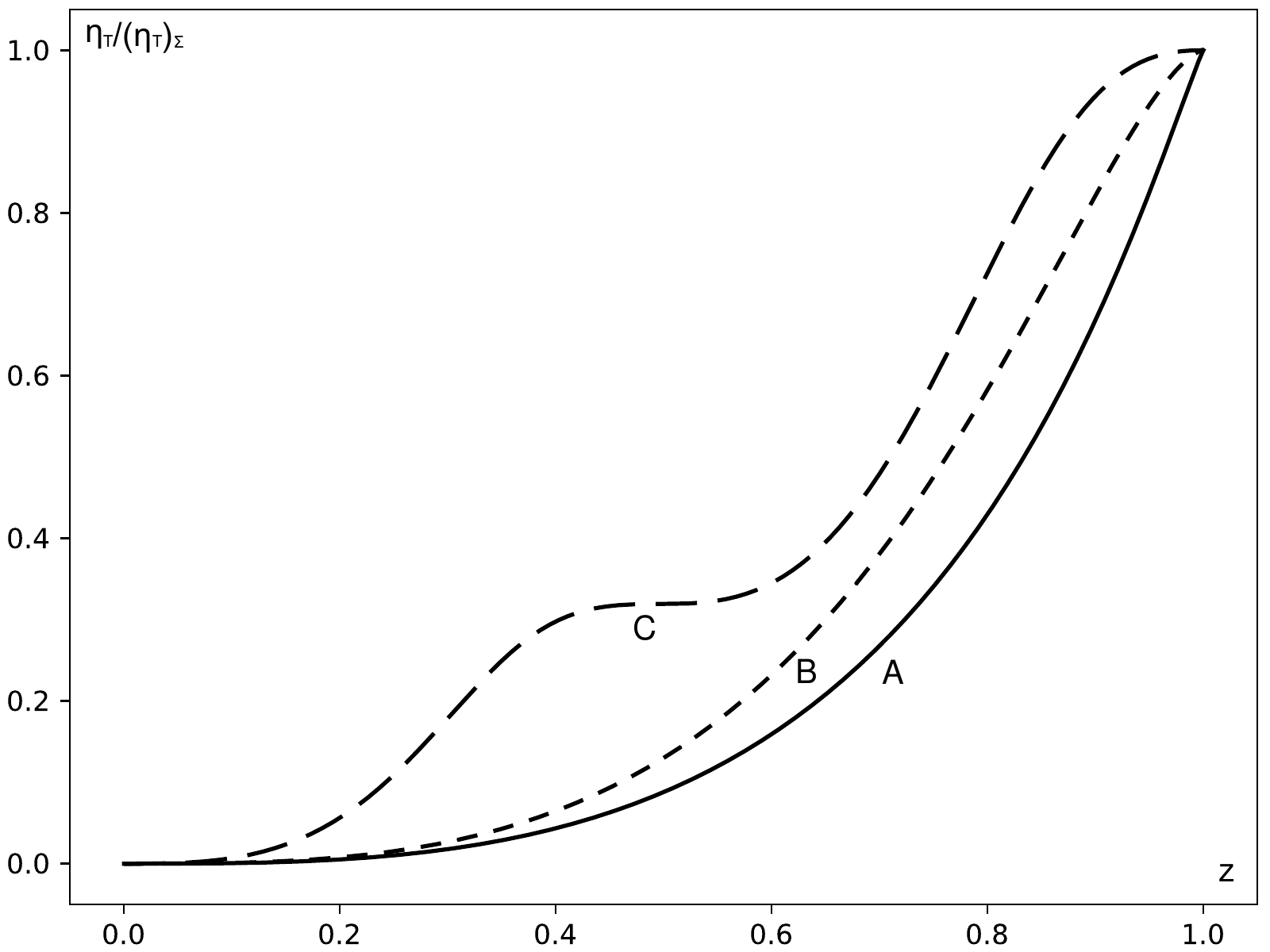}
\caption{Case 3. $\eta_T/(\eta_T)_\Sigma$ as a function of $z$ for $q_c = 0.3$ and different values of $y(n_r)$. A: $0.3728(0.1)$ (solid line), B: $0.3889(0.5)$ (short--dashed line), C: $0.2898(1.0)$ (medium--dashed line). Values of $y$ are read from Fig. \ref{comp-case03}.}
\label{TM-case03-b}
\end{figure}
The surface potential $y$ and the normalized Tolman mass, for a selection of values of the parameters are plotted in figures \ref{comp-case03} and \ref{TM-case03-b} respectively. The exhibited behavior of these variables are qualitatively the same as for a wide range of values of the parameters. Also, the conclusions extracted from  figure \ref{TM-case03-b} are basically the same as the ones reached at from the analysis of the case 1.

\section{Conclusions}
Motivated by the relevance of pressure anisotropy in the structure of self--gravitating objects, and by the fact that polytropes represent fluid systems with a wide range of applications in astrophysics (e.g. Fermi fluids),
we have described hereby a general framework for the modeling of general relativistic polytropes in the presence of anisotropic pressure, when both pressures satisfy a polytropic equation of state.  Thus, this work may be interpreted as a natural generalization of the approach described in \cite{abellan20}, to the relativistic regime. As mentioned in the Introduction, such an extension is mandatory if one has to deal with ultra compact objects  such as neutron stars, where general relativistic effects cannot be neglected.

Since the inclusion of pressure anisotropy implies an additional degree of freedom, the integration of the ensuing Lane--Emden equation, in the general case,  requires additional information. We have supplied such additional information by assuming that both pressures satisfy a polytropic equation of state. The motivation to adopt such an assumption is provided by the simple fact that for small anisotropies it is always a good approximation. For large anisotropies it is just an heuristic assumption whose validity will be confirmed (or denied) from its application to specific problems.

It should be recalled that for each polytrope there are two possible  polytropic equations of state leading to the same Newtonian limit, depending on whether we use the energy density or the baryonic mass density. Therefore for two polytropes, as is the case in this work, we have in principle four possible  cases.  We have discussed only  the  case  represented by (\ref{pr1}) and (\ref{perp1}) (i.e. we use the energy--density for both polytopes).  The treatment of the remaining  three cases is very similar and, therefore, for simplicity we have omitted here  their description.

Depending on  whether $\gamma=1$ or $\gamma\neq1$  we could differentiate three possible cases. It should be noticed that there are only three possible cases since the case $\gamma_r=\gamma_\bot$  leads to the isotropic pressure case $P_r=P_\bot$. We have integrated the Lane--Emden equations for these  three cases, for a very large set of  values of the parameters. However, only  a very specific set for each case is exhibited, since the qualitative behaviour of the system does not change much for a wide range of values of the parameters. 

Although the main reason  to present such models was not to describe any specific astrophysical scenario, but to illustrate the applications of our approach,   the obtained models exhibit some  interesting features which deserve to be commented. 

Thus we observe in figure \ref{case1_w_vs_x}  that bounded configurations exist for a range of values of  the parameters, out of which the configurations are unbounded. However due to the existence  of a larger number of parameters than in the isotropic case, the conditions for the existence of finite radius distributions are more involved than in the latter case. The same happens for the case 3, depicted in figure \ref{case3_w_vs_x}. In the case 2, instead, all configurations are unbounded, as illustrated by figure \ref{case2_w_vs_x}. This is  an expected result since this case corresponds to an isothermal gas.

Figures  \ref{TM-case01-b} and \ref{TM-case03-b} illustrate the ``strategy''  adopted by the fluid distribution   to keep the equilibrium;  it tries to concentrate  the Tolman mass in the  outer regions. This behaviour was already observed for  different families of anisotropic politropes discussed in \cite{herrera2013}. Two remarks  are in order at this point:
\begin{itemize}
\item The  ``anomalous'' behavior shown  for the innermost regions in figure \ref{TM-case01-b} is related to the extreme (maximal) value of $q_c$ that has been used for this figure ($q_c=1$), it corresponds to the stiff equation of state $P_r=\rho$, which  is believed to describe ultradense matter \cite{zh}. For smaller values of $q_c$ the above mentioned behaviour does not appear.
\item The efficiency to diminish    the Tolman mass in the inner  regions and to concentrate it in the outer ones, depends on the anisotropic factor, which  brings out  the role played by the anisotropy in the stability of the fluid configuration \cite{chan}.
\end{itemize}
Finally we would like to call the attention to the potential application of the approach presented here to the study of super-Chandrasekhar white dwarfs which may attain masses of the order of $2.8M_\odot$, and  are modeled resorting to a polytropic equation of state (see \cite{swd} and references therein). For such configurations it is evident that general relativistic effects as well as the inclusion of pressure anisotropy, are unavoidable. Nevertheless, care must be exercised with the fact that some of the physical phenomena present in such configurations (e.g. very strong magnetic fields) could break the spherical symmetry, implying thereby that our approach should be taken, in this case,  as an approximation.


\begin{thebibliography}{}
\bibitem{abellan20} G. Abell\'an, E. Fuenmayor and L. Herrera, Phys. Dark Univ. {\bf 28}, 100549  (2020).
\bibitem{LHN}L. Herera and W. Barreto, Phys. Rev D {\bf 87}, 087303 (2013).
\bibitem{1pol} S. Chandrasekhar, {\it An Introduction to the Study of Stellar Structure} (University of Chicago, Chicago, 1939).

\bibitem{3pol} R. Kippenhahn and A. Weigert, {\it Stellar Structure and Evolution} (Springer Verlag, Berlin, 1990).
\bibitem{herrera97} L. Herrera  and  N.O.  Santos,  Phys.  Rep. {\bf 286}, 53 (1997). 
\bibitem{LHP}L. Herrera, Phys. Rev. D {\bf 101}, 104024 (2020).
\bibitem{4a} R. Tooper, Astrophys. J., {\bf 140}, 434 (1964).

\bibitem{4b} R. Tooper, Astrophys. J., {\bf 142}, 1541 (1965).

\bibitem{4c} R. Tooper, Astrophys. J., {\bf 143}, 465 (1966).
\bibitem{5} S. Bludman, Astrophys. J., {\bf 183}, 637 (1973).
\bibitem{6} U. Nilsson and C. Uggla, Ann. Phys., {\bf 286}, 292 (2000).
\bibitem{7} H. Maeda, T. Harada,  H. Iguchi and N. Okuyama, Phys. Rev. D, {\bf 66}, 027501 (2002).
\bibitem{11} L. Herrera and W. Barreto, Gen. Relativ.  Gravit., {\bf 36}, 127 (2004).

\bibitem{12} X. Y. Lai and R. X. Xu, Astropart. Phys., {\bf 31},  128 (2009).

\bibitem{13} S. Thirukkanesh and F. C. Ragel, Pramana J. Phys., {\bf 78},  687 (2012).
\bibitem{herrera2013} L. Herrera and W. Barreto, Phys. Rev. D {\bf 88}, 084022 (2013).
\bibitem{prnh}L. Herrera, A. Di Prisco, W. Barreto and  J. Ospino, Gen. Relativ.  Gravit., {\bf  46}, 1827 (2014).
\bibitem{pr6}S. A. Ngubelanga and S. D. Maharaj, Eur. Phys. J. P.,{\bf 130}, 211, (2015).
\bibitem{pr9} T. Harko and M. K. Mak, Astrophys. Space Sci.,{\it 361}, 283 (2016).
\bibitem{pr10}L. Herrera, E. Fuenmayor, and P. Leon,  Phys. Rev.  D {\bf 93}, 024047  (2016).
\bibitem{pr1}H.-C. Kim, Phys. Rev., D {\it 96},  064053 (2017).
\bibitem{pr7}S. A. Mardan and M. Azam, Eur. Phys. J. C, {\bf 77}, 385 (2017).
\bibitem{pr8}M. Moussa, Ann. Phys.,{\bf 385}, 347 (2017).
\bibitem{pr3}M. Z. Bhatti and Z. Tariq, Eur. Phys. J. P. {\bf 134}, 521 ( 2019).
\bibitem{pr5} R. Roy, Pramana, {\bf 92}, 63,  (2019).
\bibitem{pr2}S. A. Mardan, A. A. Siddiqui, I. Noureen, and R. N. Jamil,  Eur. Phys. J. P., {\bf 135 }, 3 (2020).

\bibitem{pr4}M. Z. Bhatti and Z. Tariq, Phys. Dark Univ., {\bf 28},100482, (2020).

\bibitem{LHTolman1} L. Herrera, W. Barreto, A. Di. Prisco and N. O. Santos, Phys. Rev. D {\bf 65}, 104004 (2002).
\bibitem{zh} Ya. B. Zeldovich, Zh. Eksp. Teor. Fiz. {\bf 41}, 1609 (1969) [Sov. Phys. JETP  {\bf 14}, 1143 (1962)]
\bibitem{chan} R. Chan, L. Herrera and N. O. Santos, 
Month. Not. Roy. Astron. Soc. {\bf  265}, 533 (1993).
\bibitem{swd}S. Kalita, B. Mukhopadhyay, T. Mondal, and T. Bulik, {\it arXiv:2004.13750v1}[astro-ph.HE]
\end{thebibliography}
\end{document}